\documentclass[journal,10pt,twocolumn]{IEEEtran} 

\usepackage{amssymb}
\usepackage{amsthm} 
\usepackage{multirow}
\usepackage{booktabs}
\usepackage{longtable}
\usepackage{amsfonts}
\usepackage{ltablex}
\keepXColumns
\usepackage{wrapfig}
\usepackage{subcaption}
\usepackage{algorithmic,algorithm}
\usepackage{color, soul}
\ifCLASSINFOpdf
   \usepackage[pdftex]{graphicx}
  \graphicspath{{../pdf/}{../jpeg/}}
   \DeclareGraphicsExtensions{.pdf,.jpeg,.png}
\else
  \usepackage[dvips]{graphicx}
   \graphicspath{{../eps/}}
   \DeclareGraphicsExtensions{.eps}
\fi
\usepackage{cite}
\usepackage{amsmath}
\usepackage{array}
\usepackage{epstopdf}
\usepackage{extarrows}
\usepackage{hyperref}
\usepackage{multirow}
\usepackage{bm}  
\usepackage[table]{xcolor}
\usepackage{pifont}
\usepackage{supertabular}
\usepackage{scalerel}
\usepackage{float} 
\usepackage{xcolor}
\usepackage{tabularx}
\newcommand\IC{\ensuremath{\mathcal{I}}}

\newcommand\LC{\ensuremath{\mathcal{L}}}

\newcommand\SC{\ensuremath{\mathcal{S}}}

\newcommand\PB{\ensuremath{\bm{P}}}

\newcommand\wB{\ensuremath{\bm{w}}}
\newcommand\xB{\ensuremath{\bm{x}}}
\newcommand\yB{\ensuremath{\bm{y}}}

\newcommand\AF{\ensuremath{\mathbf{A}}}
\newcommand\BF{\ensuremath{\mathbf{B}}}

\newcommand\HF{\ensuremath{\mathbf{H}}}
\newcommand\FIF{\ensuremath{\mathbf{I}}}

\newcommand\VF{\ensuremath{\mathbf{V}}}

\newcommand\oneF{\ensuremath{{\mathbf 1}}}

\newcommand\CR{\ensuremath{\mathrm{C}}}
\newcommand\DR{\ensuremath{\mathrm{D}}}

\newcommand\GR{\ensuremath{\mathrm{G}}}

\newcommand\IR{\ensuremath{\mathrm{I}}}

\newcommand\LR{\ensuremath{\mathrm{L}}}
\newcommand\MR{\ensuremath{\mathrm{M}}}

\newcommand\ROR{\ensuremath{\mathrm{O}}}
\newcommand\PR{\ensuremath{\mathrm{P}}}

\newcommand\SR{\ensuremath{\mathrm{S}}}
\newcommand\TR{\ensuremath{\mathrm{T}}}

\newcommand\WR{\ensuremath{\mathrm{W}}}

\newcommand\dR{\ensuremath{\mathrm{d}}}

\newcommand\xR{\ensuremath{\mathrm{x}}}
\newcommand\yR{\ensuremath{\mathrm{y}}}

\newcommand\Rb{\ensuremath{\mathbb{R}}}

\newcommand{\wh}{\widehat}

\usepackage{acronym}
\newacro{isac}[ISAC]{integrated sensing and communication}
\newacro{rss}[RSS]{received signal strength}
\newacro{cgan}[CGAN]{conditional generative adversarial network}
\newacro{thz}[THz]{terahertz}
\newacro{sota}[SOTA]{state-of-the-art}
\newacro{mse}[MSE]{mean squared error}
\newacro{dl}[DL]{deep learning}

\newacro{pdf}[PDF]{probability density function}
\newacro{bs}[BS]{base station}
\newacro{awgn}[AWGN]{additive white Gaussian noise}
\newacro{ssb}[SSB]{synchronization
signal blocks}
\newacro{ssb}[SSB]{synchronization
signal blocks}
\newacro{dod}[DoD]{directions of departure}
\newacro{dnn}[DNN]{deep neural network}
\newacro{gan}[GAN]{generative adversarial network}
\newacro{cnn}[CNN]{convolutional neural network}

\begin{document}

\title{\LARGE Channel Gains to Captions: Task-Unified Multi-Level RF Sensing with Vision-Language Models}
\author{Tianyu~Hu,~\IEEEmembership{Student Member,~IEEE,} Zhiren~Gong,~\IEEEmembership{Student Member,~IEEE,}
        Haowei~Cui, Shuai~Wang,~\IEEEmembership{Member,~IEEE,} Samson~Lasaulce,~\IEEEmembership{Member,~IEEE,} Lingxiang~Li,~\IEEEmembership{Member,~IEEE,} Wassim~Hamidouche,~\IEEEmembership{Senior Member,~IEEE,}  
        Zhi~Chen,~\IEEEmembership{Senior Member,~IEEE,} \text{Mérouane}~Debbah,~\IEEEmembership{Fellow,~IEEE}
    \thanks{This work is supported in part by Mobile Information Networks-National Science and Technology Major Project under Grant No. 2025ZD1305200, and by the National Key R\&D Program of China under Grant 2024YFE0200400. (Corresponding authors: Lingxiang Li and Shuai Wang)}
    \thanks{Tianyu Hu, Haowei Cui, Shuai Wang, Lingxiang Li, and Zhi Chen are with the National Key Laboratory of Wireless Communications, University of Electronic Science and Technology of China (UESTC), Chengdu 611731, China (e-mail: \{huty, cchw\}@std.uestc.edu.cn; \{shuaiwang, lingxiang.li, chenzhi\}@uestc.edu.cn;).}
    \thanks{Zhiren Gong is with the Interdisciplinary Graduate Programme, Nanyang Technological University, Singapore (e-mail: zhiren001@e.ntu.edu.sg).}
    \thanks{Tianyu Hu, Samson Lasaulce, Wassim~Hamidouche, and \text{Mérouane} Debbah are with the Research Institute for Digital Future, Khalifa University, 127788 Abu Dhabi, UAE (e-mail: huty@std.uestc.edu.cn, samson.lasaulce@univ-lorraine.fr, wassim.hamidouche@insa-rennes.fr, merouane.debbah@ku.ac.ae).}
}
\IEEEaftertitletext{\vspace{-1.5\baselineskip}}
\maketitle

\begin{abstract}
This letter investigates a task-unified multi-level radio-frequency (RF) sensing framework driven by vision–language models (VLMs). Existing RF sensing methods rely on task-specific designs and provide only partial environmental information, limiting their ability to handle emerging 6G applications. To address this, we propose a generative formulation for RF sensing, where millimeter-wave (mmWave)/terahertz (THz) channel gains are mapped to captions describing multi-level environmental semantics. The framework solves this problem through a complementary design for RF-environment semantic bridging, where a VLM is fine-tuned to leverage its multimodal representations and prompt-conditioned semantic generation capabilities. Hence, different sensing tasks are specified through textual prompts, enabling the framework to handle diverse tasks in a unified manner. For fine-tuning, we introduce prompt-routed low-rank adaptation (LoRA) experts to achieve level-aware adaptation. Simulation results show that, compared with baselines, our framework achieves superior performance with a broader semantic scope, and enables task-unified sensing beyond predefined tasks. Under an unseen sensing requirement, it achieves an average F1-score improvement of $0.17$ over the most competitive variant.
\end{abstract}


\section{Introduction}
\label{sec:intro}

Integrated sensing and communication (ISAC) in millimeter-wave (mmWave)/terahertz (THz) bands is a key paradigm for sixth-generation (6G) wireless networks~\cite{10494372}. Within ISAC, radio-frequency (RF) sensing~\cite{10907867} aims to leverage RF observations to capture environmental information involved in signal propagation. Existing studies mainly focus on estimating obstacle attributes~\cite{11278185,9448728} or reconstructing physical environments~\cite{hu2025advancing,11278185}. For instance, WirelessGPT~\cite{11278185} constructed a generative foundation model based on channel state information, with task-specific heads for predefined sensing tasks. Despite their effectiveness, these methods may struggle to meet the sensing demands of emerging 6G applications (e.g., extended reality and digital twins~\cite{10494372}), which demand comprehensive environmental semantics and a unified paradigm capable of handling diverse sensing tasks, including unseen ones. This limitation stems from the reliance of existing methods on task-specific formulations and the resulting partial environmental information, leading to redesign or retraining when encountering unseen tasks.

Recently, vision–language models (VLMs)~\cite{bai2025qwen2} have shown strong potential for communication tasks~\cite{11434852,11432820} due to their knowledge-enriched multimodal representations and prompt-conditioned semantic generation capabilities. Motivated by this, we leverage VLMs to develop a framework capable of comprehensively describing environmental semantics from RF-domain observations, guided by textual prompts that specify sensing requirements. The framework is expected to possess sufficient RF-related knowledge, enabling it to exploit the VLM to generate task-specific environmental semantics, thereby operating as a unified sensing paradigm. Notably, pre-trained VLMs derive their priors from natural images and text, which are fundamentally different from RF propagation patterns, thereby lacking such knowledge and leading to an \emph{RF--VLM prior mismatch}~\cite{10892257}. To better meet the aforementioned sensing demands, the framework aligns RF observations with the semantic space through effective adaptation.


In this letter, we introduce a \emph{task-unified multi-level RF sensing framework} driven by VLMs, where we characterize the environmental semantics as \emph{multi-level semantics} spanning the layout, obstacle, and communication-related levels. Within the framework, RF sensing is cast as a generative captioning problem, where mmWave/THz channel gains are mapped to textual captions of the environment. To mitigate the prior mismatch, the framework introduces \emph{prompt-routed low-rank adaptation (LoRA) experts} to fine-tune the pre-trained VLM. These experts learn RF sensing knowledge and aggregate it using level information from the prompt, enabling level-aware adaptation while preserving the pre-trained priors. Simulation results show that, compared with baselines (e.g., [5]), the proposed framework achieves superior multi-level RF sensing performance with an expanded semantic scope, and enables task-unified sensing beyond predefined tasks. Under an unseen sensing requirement, it achieves an average F1-score improvement of $0.17$ over the most competitive variant.


\section{Preliminaries and System Model}

\subsection{Multi-Level Environmental Semantics}

\begin{figure}[t]  
  \centering
  \begin{minipage}[t]{0.44\linewidth}
    \centering
    \includegraphics[width=0.8\linewidth]{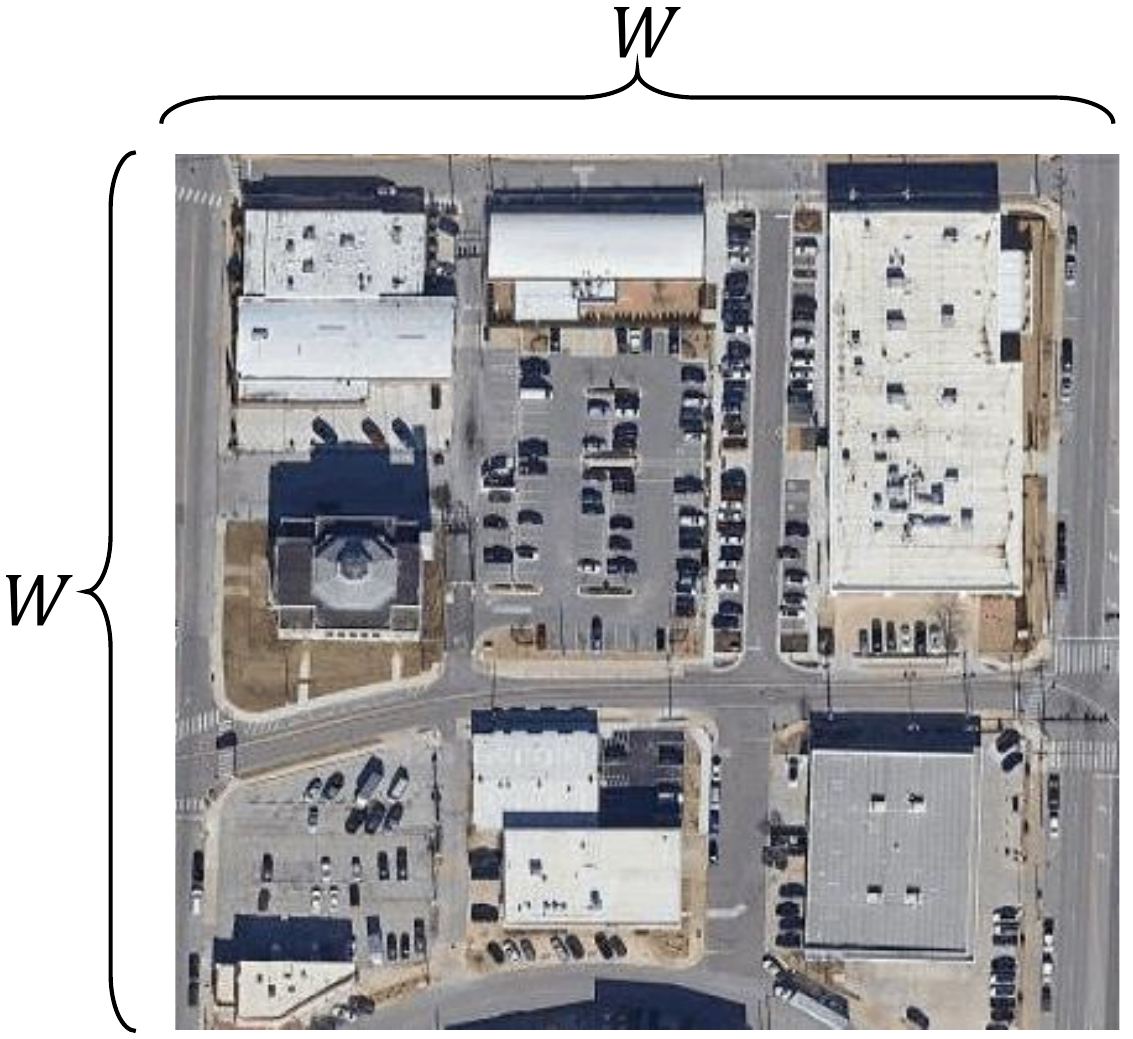}
    
    \subcaption{A section of {\it Oklahoma}.}\label{fig:a}
  \end{minipage}
  \hfill
  \begin{minipage}[t]{0.54\linewidth}
    \centering
    \includegraphics[width=0.9\linewidth]{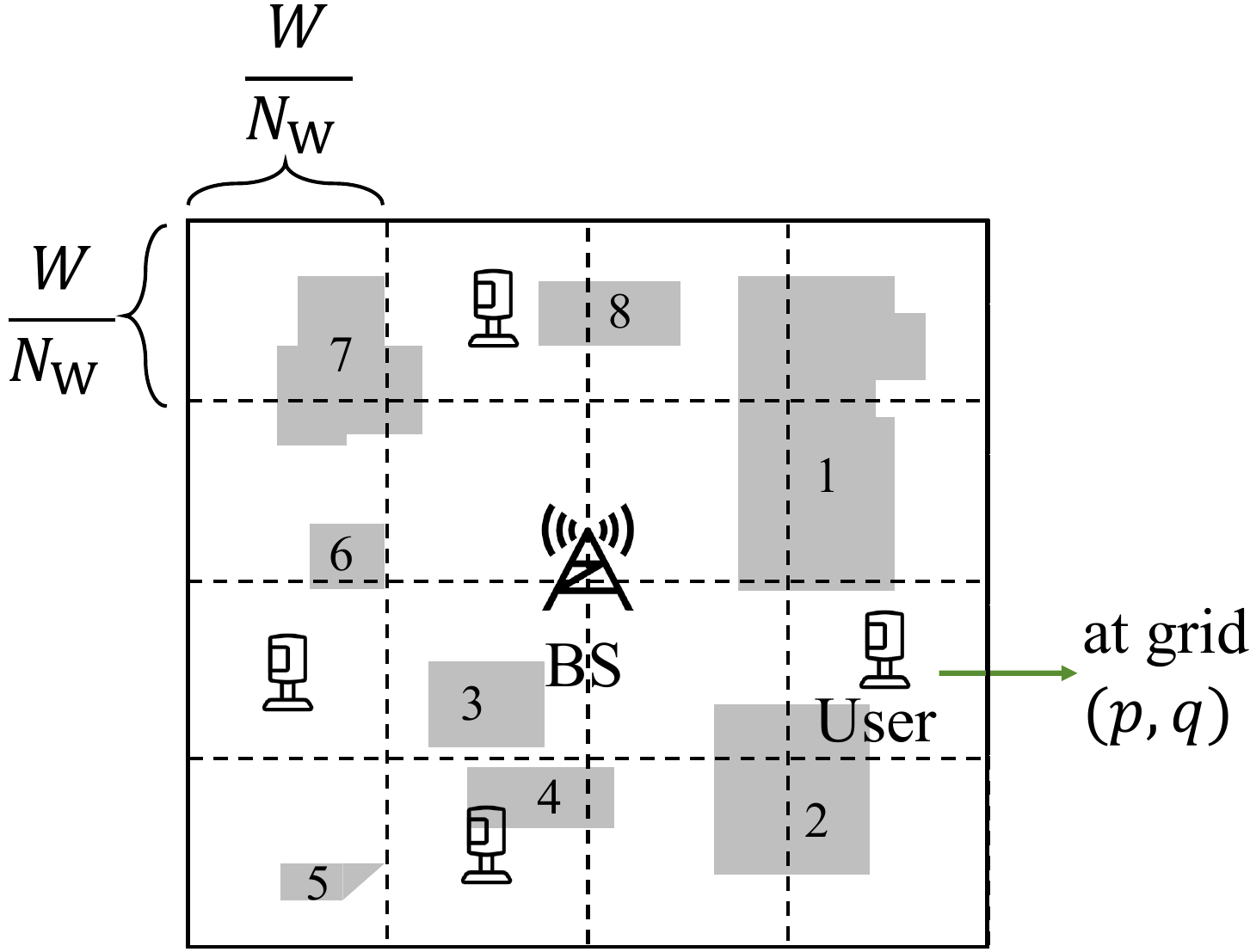}
    \subcaption{The schematic diagram.}\label{fig:b}
  \end{minipage}
  \caption{An example scenario for the considered 2D area.}
  \label{new_scenario}
\end{figure}

We consider a two-dimensional (2D) square area with side length $W$. The area contains multiple obstacles with varying shapes, sizes, and locations, leading to diverse environmental configurations. Fig.~\ref{new_scenario}(a) depicts an example scenario for the considered area, which is a satellite image of a section of {\it Oklahoma}~\cite{Alkhateeb2019}. For each scenario under this setting, we define
$\IC=\{\FIF_{\LR},\FIF_{\ROR},\FIF_{\CR}\}$ to characterize the environment at multiple semantic levels,
where $\FIF_{\LR}\in\{0,1\}^{N_{1}\times N_{2}}$, $\FIF_{\ROR}\in\{0,1\}^{N_{1}\times N_{2}\times N_{3}}$, and $\FIF_{\CR}\in\{0,1\}^{N_{1}\times N_{2}}$ represent layout, obstacle, and communication-related semantics, respectively. For all three representations, the first two dimensions index environmental attributes and their values, respectively. In addition, the third dimension in $\FIF_{\ROR}$ indexes individual obstacles. All attributes are discretized into a finite set of categories and represented by their corresponding class labels.


\textbf{Layout-level semantics ($\FIF_{\LR}$):} this level characterizes large-scale spatial properties, including the total obstacle area, as well as the spatial spread and mean of obstacle centroids. It can support communication functions involving large-scale propagation characterization.

\textbf{Obstacle-level semantics ($\FIF_{\ROR}$):} this level describes individual obstacles and their attributes, including the number of obstacles and class labels of their geometric attributes (e.g. centroid, shape, and area). It enables channel modeling and analysis for wireless links.

\textbf{Communication-related-level semantics ($\FIF_{\CR}$):} this level comprises communication-related, environment-dependent information that can be directly used for transmission configuration, including a communication-scenario label (e.g., rural, suburban, or urban) and candidate beam-direction classes with low blockage for different base-station (BS) locations.


\subsection{System Model}

We partition the considered 2D area into $N_{\WR} \times N_{\WR}$ grids. A mmWave/THz BS is located at the center of the area, and $K$ users are randomly distributed at the centers of $K$ grids, where these grid centers are not occupied by obstacles, as shown in Fig.~\ref{new_scenario}(b). To obtain the RF observations, we assume that the BS is equipped with a directional antenna and performs beam scanning over $N_{\DR}$ evenly spaced beams, while each user employs an omnidirectional antenna for reception.


For a given environment $\IC$, the channel impulse response (CIR) is capable of representing key characteristics of mmWave/THz channels, such as non-negligible molecular absorption and channel sparsity~\cite{ning2023beamforming}. Let $h_{p,q,n,\IC}(t)$ denote the CIR to a user located at the center of grid $(p,q)$ under the $n$-th beam direction, where $t$ denotes the time index. The average channel power gain $\bar{h}_{p,q,n,\IC}$ can be expressed as 
\begin{align}
\bar{h}_{p,q,n,\IC}=\frac{1}{T} \int_{0}^{T} |h_{p,q,n,\IC}(t)|^{2} \dR t,		
\label{Model_2}
\end{align}where $T$ represents the total observation time. By averaging the received signal power measured at different times within $T$, we can empirically obtain the average channel power gain $\bar{h}_{p,q,n,\IC}$. As such, channel gains collected from $K$ sparsely distributed users can be obtained. 

To provide a spatial structure, we represent the $K$ channel gains as a sparse tensor $\HF_{\IC}\in\mathbb{R}^{N_{\WR}\times N_{\WR}\times N_{\DR}}$, whose $(p,q,n)$-th entry equals $\bar{h}_{p,q,n,\IC}$ if a user is located in grid $(p,q)$, and $0$ otherwise. Since the channel gains are governed by the underlying propagation environment, $\HF_{\IC}$ can be exploited to infer $\FIF_{\LR}$ and $\FIF_{\ROR}$, thereby enabling the further inference of $\FIF_{\CR}$. As a result, the channel gains can be regarded as a natural basis for multi-level RF sensing.

\begin{figure*}[!t]
\centering
\includegraphics[width=1\linewidth] {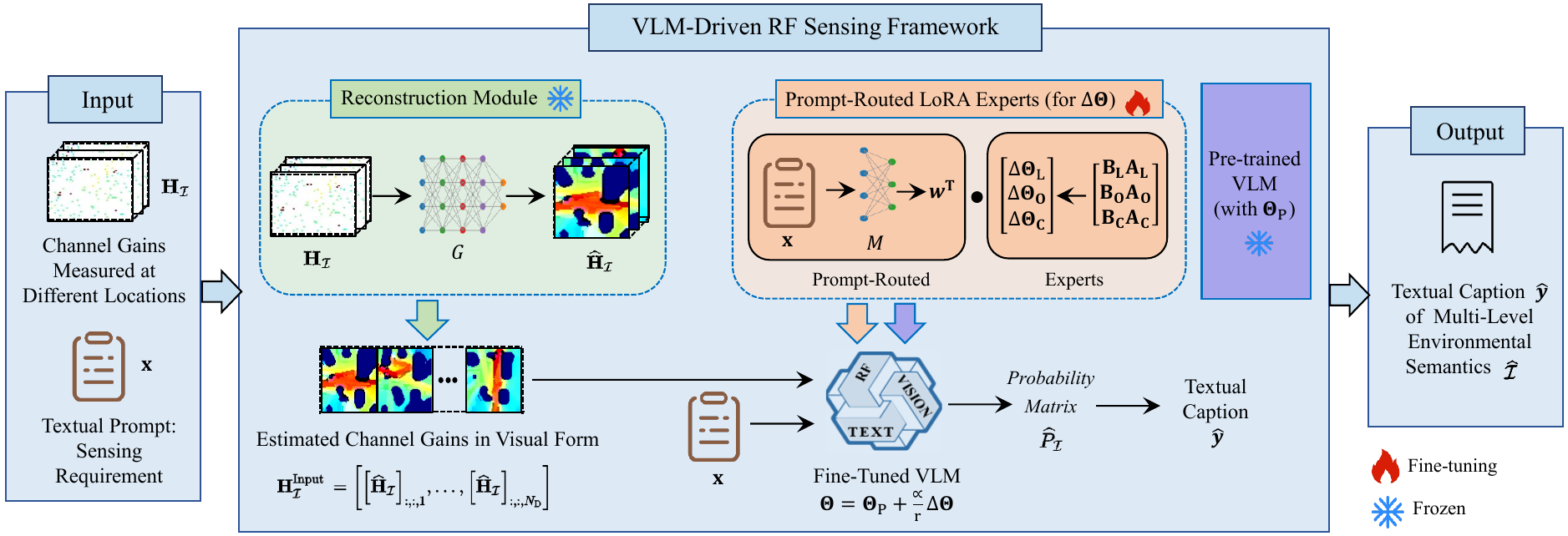} 

\caption{\centering The proposed VLM-driven RF sensing framework.}
\label{overall_framework}
\end{figure*}

\section{VLM-Driven RF Sensing Framework}

In this section, we propose a task-unified multi-level RF sensing framework driven by VLMs, which captures the environmental semantics $\IC$ from channel gains $\HF_{\IC}$ conditioned on sensing task requirements, as illustrated in Fig.~\ref{overall_framework}. 

\subsection{Problem Formulation}

Motivated by the prompt-conditioned semantic generation capability of VLMs, the proposed framework casts the RF sensing problem as a captioning task, thereby enabling a unified paradigm for diverse sensing tasks. Specifically, the captioning task introduces textual prompts to specify sensing requirements, and generates corresponding environmental captions from RF-domain channel gains. Such a generative formulation not only allows sensing requirements to vary flexibly, but also renders the output space compositional, enabling captions with flexible length, content richness, and semantic levels. In contrast, discriminative formulations adopted by existing methods are restricted to specific sensing requirements and impose a fixed output space.


Given a pre-defined token vocabulary $\VF$ and a textual prompt represented as a token sequence $\xB\in\VF^{N_{\xR}}$, we define a probability generation function $f\!:\! \Rb^{N_{\WR}\times N_{\WR}\times N_{\DR}} \!\times\! \VF^{N_{\xR}}\!\!\to\!\! [0,1]^{N_{\yR}'\times|\VF|}$ that outputs a probability matrix $\wh{\PB}_{\IC}=f(\HF_{\IC},\xB)$ for the environment with $\IC$. Each row of $\wh{\PB}_{\IC}$ corresponds to a token position and forms a probability distribution over $\VF$.  By maximum-probability decoding from $\wh{\PB}_{\IC}$, we can obtain a token sequence $\wh{\yB}\in\VF^{N_{\yR}'}$, i.e., the generated environmental caption. In other words, the proposed framework generates captions via a probability distribution over $\VF$, where the $j$-th row $[\wh{\PB}_{\IC}]_{j,:}$ corresponds to the token $[\wh{\yB}]_{j}$. 


Assuming a ground-truth caption $\yB\in\VF^{N_{\yR}}$ of environmental semantics $\IC$\footnote{For example, an instance of $\IC$ can be described as “this scene contains seven obstacles, each of which is rectangular,” and then tokenized into $\yB$.}, with a corresponding row-wise one-hot matrix $\PB_{\IC}\in\{0,1\}^{N_{\yR}\times|\VF|}$, the captioning problem for RF sensing is formulated as follows:
\begin{subequations}\label{construction}
\begin{align}
	\underset{f}{\mathrm{minimize}}~~&\mathbb{E}_{\IC}\!\left[\LC\left(\wh{\PB}_{\IC},\PB_{\IC}\right)\right], \label{const:2a} \\
	\text{s.t.}~~& \wh{\PB}_{\IC}=f(\HF_{\IC},\xB), \label{const:2d}\\
	& \wh{\PB}_{\IC}\oneF=\oneF, \label{const:2f}\\
	& \PB_{\IC}=g(\IC). \label{const:2b}
\end{align}
\end{subequations} The cross-entropy loss $\LC(\wh{\PB}_{\IC},\PB_{\IC})$ is employed, as it corresponds to maximum likelihood estimation of the token distribution and effectively penalizes deviations from the ground-truth one-hot targets. The mapping $g(\IC)=g_1(g_2(\IC))$ consists of: $g_2(\IC)$ verbalizing $\IC$ into a narrative token sequence $\yB$, and $g_1(\yB)$ converting $\yB$ into $\PB_{\IC}$ by assigning each token in $\yB$ to a one-hot vector over $\VF$. Accordingly, RF sensing is formulated as an ontology-grounded, prompt-conditioned RF caption generation problem, where more diverse semantic representations and more flexible linguistic descriptions can be incorporated by modifying $\mathcal{I}$ and $g_2(\mathcal{I})$.

At the same time, solving problem~\eqref{construction} remains challenging due to the following issues. First, the limited RF prior information, together with the high dimensionality of the obstacle attribute space, may result in an excessively large feasible solution space. Second, a substantial level-dependent gap exists between RF representations and the multi-level linguistic semantics of the environment. Different semantic levels rely on distinct aspects of prior RF knowledge, making a single adaptation path prone to suboptimal alignment for some levels. To mitigate the above challenges and solve problem~\eqref{construction}, we carefully design the framework architecture and fine-tuning strategy, as detailed in the following subsections.

\subsection{Framework Architecture}

The proposed framework consists of a pre-trained VLM, a reconstruction module, and three prompt-routed LoRA experts corresponding to the layout, obstacle, and communication-related levels. The framework takes the channel gains $\HF_{\IC}$ and textual prompt $\xB$ as input, and first implements the function $f$ to produce $\wh{\PB}_{\IC}$. Then, by maximum-probability decoding, we obtain the textual caption $\wh{\yB}$, which is embedded with the sensed multi-level environmental semantics $\wh{\IC}=\{\wh{\FIF}_{\LR},\wh{\FIF}_{\ROR},\wh{\FIF}_{\CR}\}$. By leveraging the VLM’s  embedding matrix to compute semantic similarity, $\wh{\IC}$ can be parsed from $\wh{\yB}$. 


Pre-trained VLM learns powerful visual and textual representations enriched with extensive semantic knowledge. Such properties make VLM well-suited as the backbone of the proposed framework for alleviating the aforementioned challenges in problem~\eqref{construction}. Specifically, their implicit knowledge of real-world spatial structures can be leveraged to constrain the feasible solution space. Meanwhile, the RF-domain channel gains $\HF_{\IC}$ can be represented as images and fed into the VLM, thereby enabling the extraction of informative features.

The reconstruction module serves to enhance RF priors and further reduce the feasible solution space under sparse observations $\HF_{\IC}$. This module employs a pre-trained network $G$ with parameters $\theta_{\GR}$~\cite{hu2025advancing} to transform $\HF_{\IC}$ into a complete tensor, i.e., $\wh{\HF}_{\IC}=G(\HF_{\IC};\theta_{\GR})$, which provides auxiliary channel gain estimates over unmeasured grids. Then, we concatenate the $N_{\DR}$ slices of $\wh{\HF}_{\IC}$ and transform them into an image, yielding the estimated channel gains in visual form, denoted by $\HF_{\IC}^{\mathrm{Input}}=\big[[\wh{\HF}_{\IC}]_{:,:,1},\ldots,[\wh{\HF}_{\IC}]_{:,:,N_{\DR}}\big]\in\mathbb{R}^{N_{\WR}\times N_{\DR}N_{\WR}}$, which is used as the visual input to the VLM.

To address the level-dependent RF–VLM prior mismatch in a pre-trained VLM, we propose prompt-routed LoRA experts within the framework. This module leverages the level information contained in the prompt $\xB$ to perform a weighted combination of RF sensing knowledge from three LoRA experts at their respective semantic levels. Specifically, the LoRA experts are parameterized by $\Delta\mathbf{\Theta}'=[\Delta\mathbf{\Theta}_{\LR},\Delta\mathbf{\Theta}_{\ROR},\Delta\mathbf{\Theta}_{\CR}]^{\TR}$, where RF sensing knowledge is learned via LoRA-based fine-tuning (see the following subsection) and implicitly encoded in these parameters. A weight vector $\wB$ is generated by a two-layer neural network $M$ with parameters $\theta_{\MR}$~\cite{liu2024moe}, taking the embedding $\xB_{\mathrm{e}}$ of the prompt $\xB$ as input\footnote{The embedding $\xB_{\mathrm{e}}$ is obtained by mapping the prompt $\xB$ through a pre-trained embedding matrix~\cite{bai2025qwen2}.}. As a result, the parameter update $\Delta\mathbf{\Theta}=M(\xB_{\mathrm{e}};\theta_{\MR})^{\TR}\Delta\mathbf{\Theta}'=\wB^{\TR}\Delta\mathbf{\Theta}'$.  

With the parameter update  $\Delta\mathbf{\Theta}$ and the pre-trained VLM parameters $\mathbf{\Theta}_{\PR}$, the parameters of the fine-tuned VLM are given by $\mathbf{\Theta}=\mathbf{\Theta}_{\PR}+\frac{\alpha}{r}\Delta\mathbf{\Theta}$, where $\alpha$ and $r$ are the scaling factor and rank used in LoRA-based fine-tuning, respectively.

\subsection{LoRA-based Fine-Tuning}

LoRA-based fine-tuning exploits the low “intrinsic rank”~\cite{hu2022lora} of weight updates to enable parameter-efficient adaptation, which helps bridge the level-dependent gap. For each LoRA expert, $\Delta\mathbf{\Theta}_{i\in \{\LR, \ROR, \CR\}}$ is realized by introducing a low-rank update $\BF_{i}\AF_{i}$, 
which is zero-padded into the full parameter space of the VLM. 
Here, $\BF_{i}\in\mathbb{R}^{a\times \frac{r}{3}}$, $\AF_{i}\in\mathbb{R}^{\frac{r}{3}\times b}$, and $r\ll \min(a,b)$. The optimization of $\BF_{i}$ and $\AF_{i}$, i.e., the LoRA-based fine-tuning of the pre-trained VLM, is carried out using the following empirical objective, 
which is a reformulation of problem~\eqref{construction}:
\begin{subequations}\label{final_construction}
\begin{align}
\underset{\BF',\AF',\theta_{\MR}}{\mathrm{minimize}}~~&
\frac{1}{N_{\SR}}\sum_{s=1}^{N_{\SR}}\LC\!\left(\wh{\PB}_{s,\IC},\PB_{s,\IC}\right) \label{const:3a}\\
\text{s.t.}~~&
[\wh{\PB}_{s}]_{j,:}=f_{\mathbf{\Theta}}\!\left([\wh{\yB}_{s}]_{j}\mid [\wh{\yB}_{s}]_{<j},\HF^{\mathrm{Input}}_{s,\IC},\xB\right), \label{const:3d}\\
&
\mathbf{\Theta}\!=\!\mathbf{\Theta}_{\PR}+\frac{\alpha}{r}M(\xB_{\mathrm{e}};\theta_{\MR})^{\TR}d(\BF',\AF'), \label{const:3r}
\end{align}
\end{subequations} where $\BF'=[\BF_{\LR},\BF_{\ROR},\BF_{\CR}]^{\TR}$, $\AF'=[\AF_{\LR},\AF_{\ROR},\AF_{\CR}]^{\TR}$, $j\in\{1,\ldots,N_{\yR}'\}$, and $d(\BF',\AF')$ maps the low-rank matrices in $\BF'$ and $\AF'$ 
to the parameter updates $\Delta\mathbf{\Theta}'$. $N_{\SR}$ is the number of fine-tuning samples. Notably, constraint~\eqref{const:3d} reflects the autoregressive mechanism of the VLM~\cite{bai2025qwen2}.


\section{Experimental Results}
\label{sec:res}


{\bf Datasets and Model Hyperparameters.} We consider three types of 2D scenarios~\cite{hu2025advancing}, which include fine-tuning and validation scenarios with simulated environment, as well as validation scenarios based on the real-world city section shown in Fig.~\ref{new_scenario}(b). These scenarios are labeled as \( \SC_1 \), \( \SC_2 \), and \( \SC_3 \), respectively. When a model is fine-tuned on \( \SC_1 \), scenarios \( \SC_2 \) and \( \SC_3 \) are unseen with respect to that model. Specifically, $\SC_1$ contains $250$ scenarios, each generating $7$ samples with prompts specifying all combinations of semantic levels. The number of obstacles in $\SC_1$ ranges from ${1,2,3,4,5}$, with $50$ scenarios for each case. $\SC_2$ contains $50$ scenarios, each with $6$ obstacles. $\SC_3$ contains $8$ scenarios, constructed by sequentially removing obstacles according to their indices in Fig.~\ref{new_scenario}(b). Each scenario in \( \SC_1\) and \( \SC_2\) contains obstacles with randomly generated quadrilateral shapes and positions. We perform ray tracing to obtain mmWave/THz channel gain observations. For the communication-related-level information $\FIF_{\CR}$, we consider $4$ predefined BS locations at the centers of the four quadrants. Note that we use Qwen2.5-VL~\cite{bai2025qwen2} as the pre-trained VLM.

{\bf Baseline Models and Evaluation Metrics.} 
Existing methods are typically tailored to specific sensing tasks and therefore cannot support task-unified and multi-level RF sensing. As a compromise, we include U-NetGAN~\cite{hu2025advancing}, ResNet-18~\cite{7780459}, and ablated variants of the proposed framework as baselines. U-NetGAN is task-specific and can only infer $\wh{\FIF}_{\LR}$ and $\wh{\FIF}_{\ROR}$ via post-processing of segmented instances within sensed environmental maps, but cannot infer $\wh{\FIF}_{\CR}$. ResNet-18 enables multi-level RF sensing via classification, but its fixed output space prevents task-unified capability. We adopt the averaged macro-F1 score~\cite{11200914} to evaluate the sensed information $\wh{\IC}=\{\wh{\FIF}_{\LR},\wh{\FIF}_{\ROR},\wh{\FIF}_{\CR}\}$, where a higher F1 score indicates better sensing performance in terms of precision and recall. For $\wh{\FIF}_{\LR}$ and $\wh{\FIF}_{\ROR}$, whose attributes are represented in a one-hot manner, the F1 score reduces to the inner product, i.e., $\frac{1}{N_{1}N_{\SR}}\langle\operatorname{vec}(\FIF_{\LR}), \operatorname{vec}(\wh{\FIF}_{\LR})\rangle$ and $\frac{1}{N_{1}N_{3}N_{\SR}}\langle\operatorname{vec}(\FIF_{\ROR}), \operatorname{vec}(\wh{\FIF}_{\ROR})\rangle$, where $\operatorname{vec}(\cdot)$ denotes the vectorization operation. For $\wh{\FIF}_{\CR}$, the F1 score is given by $\frac{1}{N_{1}N_{\SR}}\sum_{s}^{N_{\SR}}\sum_{c}^{N_{1}}\frac{2\langle [\wh{\FIF}_{\CR}]_{c,:},[\FIF_{\CR}]_{c,:}  \rangle}{\left\|[\wh{\FIF}_{\CR}]_{c,:}\right\|_1+\left\|[\FIF_{\CR}]_{c,:}\right\|_1}$. All reported results are averaged over five independent runs, each using an independently trained reconstruction module and a different user-location configuration.


\renewcommand{\arraystretch}{1.3}   
\setlength{\tabcolsep}{5pt}          

\begin{table}[htbp]
\centering
\caption{Scenario Details and Model Hyperparameters.}
\label{tab_evalution}

\begin{tabular}{
  >{\centering\arraybackslash}p{0.18\textwidth}|
  >{\centering\arraybackslash}p{0.27\textwidth}}\hline
\textbf{Description} & \textbf{Value} \\\hline
Area size & $W\times W=100\,\mathrm{m}\times100\,\mathrm{m}$ \\\hline
Grid division & $N_{\WR}\times N_{\WR}=64\times64$ \\\hline
Carrier frequency & $300\,\mathrm{GHz}$ \\\hline
Number of beams and users & $N_{\DR}=18$ and  $K=150$\\\hline
Learning rate & $0.0001$ \\\hline
Scaling factor and rank & $\alpha=48$ and $r=24$ \\\hline
\end{tabular}
\end{table}

\begin{figure}[!t]
\center
\includegraphics[width=\linewidth] {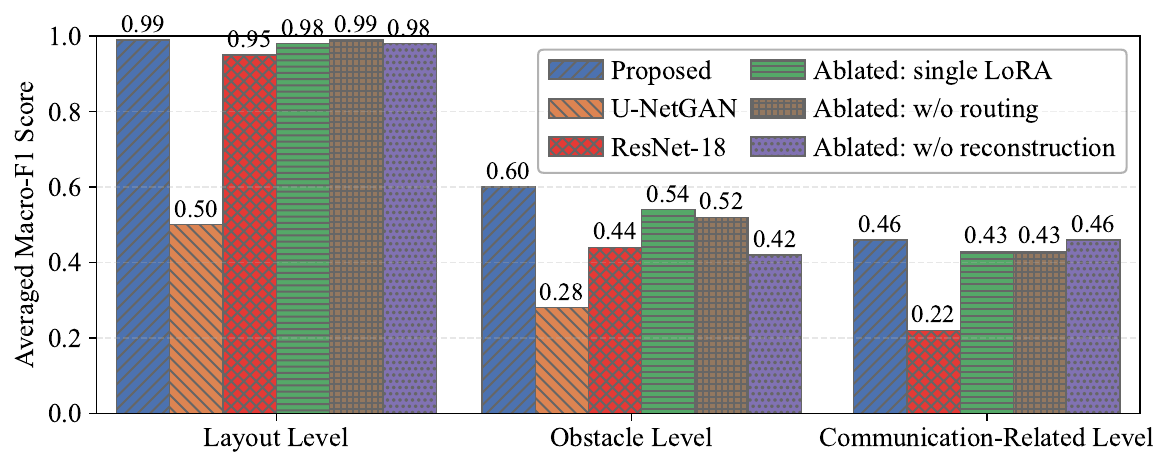}
\caption{Averaged macro-F1 scores for multi-level environmental semantics sensed by different methods in $\SC_2$.}
\label{ALL_obstacle_RFOB}
\end{figure}

\begin{figure*}[!t]
  \centering
\includegraphics[width=0.9\linewidth]{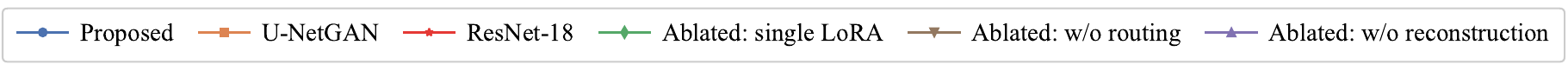}
  \vspace{0.3em}
  \begin{subfigure}[b]{0.30\textwidth}
\includegraphics[width=0.9\linewidth]{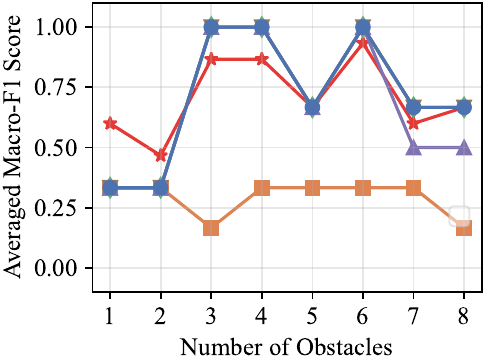}
    \caption{Layout Level}
    \label{all_GT}
  \end{subfigure}
  \hfill
  \begin{subfigure}[b]{0.30\textwidth}
\includegraphics[width=0.9\linewidth]{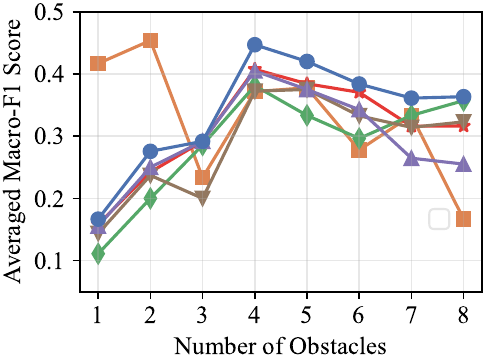}
    \label{all_231}
    \caption{Obstacle Level}
  \end{subfigure}
  \hfill
  \begin{subfigure}[b]{0.30\textwidth}
\includegraphics[width=0.9\linewidth]{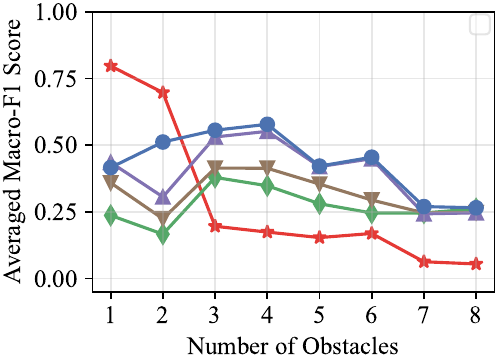}
    \label{all_236}
    \caption{Communication-Related Level}
  \end{subfigure}
  \caption{Averaged macro-F1 scores versus the number of obstacles for different methods in $\SC_3$.}
  \label{ALL_obstacle}
\end{figure*}


Fig.~\ref{ALL_obstacle_RFOB} reports the averaged macro-F1 scores for multi-level environmental semantics sensed by different methods in $\SC_2$. The proposed framework generally achieves superior performance over all baselines across the layout, obstacle, and communication-related levels. Nevertheless, it achieves only moderate sensing capability for obstacle attributes and communication-related semantics. Moreover, the ablation studies in Fig.~\ref{ALL_obstacle_RFOB} demonstrate that the effectiveness of the framework is attributed not only to the RF sensing knowledge enabled by fine-tuning, but also to the prompt routing and the reconstruction module, which respectively enable adaptive knowledge fusion and provide enhanced RF priors.

Under the scenarios \( \SC_{3} \), we investigate the impact of the number \( N_{\IR} \) of obstacles on the proposed framework and the baselines, as shown in Fig.~\ref{ALL_obstacle}. The proposed framework achieves higher F1 scores than the baselines in more complex scenarios, i.e., when  \( N_{\IR}>2 \). This is likely because more obstacles introduce richer signal-environment interactions and provide more sensing cues, allowing the proposed framework to better exploit its learned semantic understanding and reasoning capability. The proposed framework and its ablated variants generally exhibit an increasing-then-decreasing trend as the number of obstacles grows, indicating that the benefits brought by additional cues gradually saturate, while overly complex environments make RF sensing more challenging.

In Table~\ref{tab_integration_new}, we compare the F1 scores of different methods in $\SC_3$ on an unseen sensing requirement not involved during fine-tuning, where the VLM is prompted to infer possible beam directions for a BS at the scenario center. As shown in Table~\ref{tab_integration_new}, the proposed framework consistently outperforms the ablated variants across different obstacle numbers, with an average improvement of $0.17$ in F1 score over the most competitive variant (i.e., ``Ablated: w/o routing"). In Table~\ref{tab_integration_new}, the proposed framework achieves an average relative improvement of $64\%$ over the ablated variants, substantially exceeding the corresponding $8\%$ and $13\%$ improvements observed in Fig.~\ref{ALL_obstacle_RFOB} and Fig.~\ref{ALL_obstacle}, respectively. This result further highlights the importance of our complementary design for addressing unseen sensing requirements. It should be noted that the proposed framework achieves an F1 score of only $0.09$ in the eight-obstacle case, indicating its limited sensing capability under unseen and highly complex conditions.

\renewcommand{\arraystretch}{1}   
\setlength{\tabcolsep}{6pt} 
\begin{table}[ht]
\centering
\caption{F1 Score Comparison on an Unseen Task.}
\label{tab_integration_new}

\begin{tabular}{|
  >{\centering\arraybackslash}m{0.10\textwidth}|
  >{\centering\arraybackslash}m{0.020\textwidth}|
  >{\centering\arraybackslash}m{0.020\textwidth}|
  >{\centering\arraybackslash}m{0.020\textwidth}|
  >{\centering\arraybackslash}m{0.020\textwidth}|
  >{\centering\arraybackslash}m{0.020\textwidth}|
  >{\centering\arraybackslash}m{0.020\textwidth}|
  >{\centering\arraybackslash}m{0.020\textwidth}|
  >  {\centering\arraybackslash}m{0.020\textwidth}|}
\hline

\multirow[c]{2}{*}{\textbf{Method}} 
& \multicolumn{8}{c|}{\textbf{Number of obstacles in $\SC_3$}} \\ \cline{2-9}
& \textbf{1} & \textbf{2} & \textbf{3} & \textbf{4} & \textbf{5} & \textbf{6} & \textbf{7} & \textbf{8} \\
\hline

\textbf{Proposed} & 0.68 & 0.38 & 0.38 & 0.34 & 0.38 & 0.23 & 0.22 & 0.09 \\\hline

\textbf{Ablated: single LoRA} & 0.25 & 0.17 & 0.13 & 0.13 & 0.10 & 0.06 & 0.04 & 0.08 \\\hline

\textbf{Ablated: w/o routing} & 0.40 & 0.23 & 0.12 & 0.15 & 0.15 & 0.11 & 0.13 & 0.05 \\\hline

\textbf{Ablated: w/o reconstruction} & 0.12 & 0.13 & 0.09 & 0.07 & 0.06 & 0.03 & 0.02 & 0.03 \\\hline
\end{tabular}
\end{table}

\section{Conclusion}

In this letter, we have proposed a novel framework that adapts VLMs for task-unified multi-level RF sensing. The framework has been designed to process mmWave/THz channel gains and diverse sensing requirements, generating environmental semantic captions across the layout, obstacle, and communication-related levels. Simulation results show that, compared with baselines, the proposed framework achieves superior RF sensing performance and can generalize to diverse sensing requirements.

\bibliographystyle{IEEEtran}
\bibliography{ref}

\end{document}